%% file: main.tex
\RequirePackage[2024-06-01]{latexrelease}
\documentclass[aip,jcp,reprint,amsmath,amssymb,floatfix]{revtex4-2}
\usepackage[T1]{fontenc}
\usepackage[utf8]{inputenc}
\usepackage{lmodern}
\usepackage{graphicx}
\usepackage{booktabs}
\usepackage{array}[=2023-11-01]
\usepackage{tabularx}
\usepackage{hyperref}
\hypersetup{hidelinks,pdftitle={Perspective: The HSE Screened Hybrid and the Band Gap Problem: Origins, Impact, and the Contenders},pdfauthor={Gustavo E. Scuseria}}
\newcolumntype{L}[1]{>{\raggedright\arraybackslash}p{#1}}
\newcolumntype{C}[1]{>{\centering\arraybackslash}p{#1}}
\newcolumntype{Y}{>{\raggedright\arraybackslash}X}
\newcolumntype{Z}{>{\centering\arraybackslash}X}

\makeatletter
\let\auto@bib@innerbib\@empty
\let\write@bibliographystyle\relax
\makeatother
\begin{document}
\title{Perspective: The HSE Screened Hybrid and the Band Gap Problem: Origins, Impact, and the Contenders}
\author{Gustavo E. Scuseria}
\affiliation{Departments of Chemistry, Physics and Astronomy, and Materials Science and NanoEngineering,\\
Rice University, Houston, Texas 77005, USA}
\begin{abstract}
The ideas behind the screened hybrid functional HSE took shape in the early 2000s, about a quarter century ago; the functional appeared in this journal in 2003. HSE was not explicitly designed to predict band gaps, although at the time we did recognize that hybrids could help address the band-gap problem. It was designed to make exact exchange affordable and physically defensible in extended systems. Keeping exact exchange at short range and screening it away at long range inverted the prescription developed for molecules. When HSE band gaps for ordinary semiconductors and selected metal oxides came out substantially closer to experiment than those obtained with conventional semilocal approximations, we were pleasantly surprised and encouraged by the path we had taken. The revision from HSE03 to HSE06 improved molecular thermochemistry while preserving the good accuracy for semiconductor gaps. The one-quarter exchange fraction was inherited unchanged from the successful unscreened PBE hybrid for molecules, rather than fitted to solids. HSE established a practical screened-hybrid approach for accurate band-gap calculations, expanding first-principles work in defect physics, photovoltaics, power electronics, and quantum information, among other areas. The first part of this Perspective recounts the decisions made at HSE's birth and their influence on subsequent developments. The second part assesses recent advances in semilocal and hybrid functionals against benchmarks on sets of solids. Recent meta-GGAs achieve HSE06-level accuracy for selected semiconductors, while dielectric-dependent and tuned hybrids address the limitations of fixed screening. These hybrid developments share with our early work on local hybrids and local range separation the aim of adapting exact exchange to the electronic environment. The comparisons assess band gaps alongside structures, energetics, computational cost, and transferability.
\end{abstract}
\maketitle
\enlargethispage{\baselineskip}

% Word vs9 paragraph 006
\section{Introduction}\label{sec:introduction}

% Word vs9 paragraph 007
The band gap of a semiconductor or insulator is among the most consequential quantities in materials modeling and among the most awkward to compute. The awkwardness is structural rather than technical. The Kohn--Sham eigenvalue gap of the exact functional is not the fundamental gap; the two differ by the derivative discontinuity of the exchange-correlation potential at integer particle number.\cite{ref001,ref002} Conventional local and semilocal approximations applied in the pure Kohn--Sham scheme, where the potential is multiplicative and has no discontinuity, underestimate the gaps of semiconductors and insulators substantially and systematically. In PBE, silicon comes out near 0.6 eV against an experimental 1.17 eV, GaAs near 0.4 eV against 1.52 eV, and ZnO near 0.7 eV against 3.4 eV; reported mean absolute relative errors of PBE gaps are roughly 45--60\%, depending on the benchmark set.\cite{ref003,ref004} For decades, this limitation has stood in the way of quantitative gap prediction for the materials of greatest concern in applications.

% Word vs9 paragraph 008
The generalized Kohn--Sham (GKS) framework provides the escape.\cite{ref005} When the exchange-correlation potential is a nonmultiplicative operator, part of the discontinuity is transferred into the eigenvalue spectrum. Under stated conditions --- a continuous GKS operator and a delocalized added or removed density in an extended system --- the GKS gap equals the fundamental gap of the approximate functional in question.\cite{ref006,ref007} This is a statement about the approximation, not a guarantee of agreement with the exact physical gap; how good the resulting number is remains an empirical matter. Hybrid functionals realize the nonmultiplicative potential through explicit nonlocal exchange, and orbital-dependent meta-GGAs realize it through the kinetic energy density. The relationship between those two routes is the subject of the second part of this Perspective.

% Word vs9 paragraph 009
The first part is history, and its occasion is an anniversary. The ideas behind the Heyd--Scuseria--Ernzerhof screened hybrid took shape in the early 2000s, about a quarter century ago, and the paper introducing it appeared in this journal in 2003.\cite{ref008} This anniversary offers an occasion to set down how the functional came about and to take stock of band-gap prediction today. Earlier screened-exchange schemes provided important precedents.\cite{ref009} HSE brought together a specific functional construction, an efficient implementation, and a program of validation that established its usefulness for molecules and solids.\cite{ref008,heyd2004molecular,ref037,ref003} I have described the broader personal history in a scientific autobiography and discussed later advances in a commentary on band-gap prediction.\cite{ref010,audit510} Here I revisit the decisions that shaped HSE, and then assess what subsequent developments reveal about their strengths and limitations. A retrospective concluding that its subject was permanent would be worth nobody's time. The narrower claim advanced here is that the distance dependence of exact exchange in an extended system is an important design variable, whose consequences the field is still working out, and that the influence of the ideas behind HSE extends beyond the particular parameterization which first carried it into general use.

% Word vs9 paragraph 010
\section*{\texorpdfstring{\makebox[\columnwidth]{Origins and Impact}}{Origins and Impact}}

% Word vs9 paragraph 011
\section{Where HSE Came From}\label{sec:origins}

% Word vs9 paragraph 012
\subsection{Infrastructure first}\label{sec:infrastructure}

% Word vs9 paragraph 013
HSE grew out of a decade of work in our group to remove the principal computational bottlenecks in applying Gaussian-orbital electronic-structure calculations to large systems. We developed linear-scaling treatments of the Coulomb interaction and exchange-correlation quadrature, together with a near-linear-scaling Hartree--Fock exchange algorithm for insulating systems.\cite{ref011,stratmann1996quadratures,burant1996exchange} Conjugate-gradient density-matrix search addressed the cubic-scaling diagonalization bottleneck when sufficient sparsity could be exploited.\cite{millam1997density} Kudin extended fast multipole methods to periodic lattices with arbitrary unit-cell geometries and to periodic Gaussian charge distributions.\cite{kudin1998cells,ref014} These developments, reviewed in 1999, underpinned our periodic Gaussian-orbital implementation of Hartree-Fock and density-functional energies, as well as their analytic forces.\cite{scuseria1999linear,ref015} The broader program also reached second-order perturbation theory and coupled cluster theory under appropriate locality conditions.\cite{ref012,ref013} Konstantin Kudin steered the periodic code, and Artur Izmaylov later improved it, bringing HSE costs close to nonhybrid DFT in the systems tested.\cite{audit281} This infrastructure supported the original periodic implementation in Gaussian\cite{ref016} and the early HSE calculations on solids.\cite{ref008,ref037,ref003}

The distinction between Coulomb and exchange costs was central. Fast multipole methods accelerated the electrostatic problem for the total charge density; applying the same idea directly to exact exchange would still leave many orbital-pair contributions to evaluate.\cite{ref014,ref015,scuseria1999linear} Exchange, therefore, required its own screening and locality arguments, even though near-linear scaling was already possible for insulating molecular systems.\cite{burant1996exchange} The spatial decay of the one-particle density matrix is decisive: it is exponential in insulators, with a rate related to the gap and electronic structure, whereas at zero temperature its decay in metals is algebraic; finite temperature introduces exponential damping.\cite{ref017,ref018,ref019} Retaining only short-range Fock exchange offered a way to reduce the periodic exchange cost substantially. The remaining question was whether that computational simplification would preserve the physical advantages of a hybrid.

\subsection{A global hybrid for solids}\label{sec:global}

Before HSE, there was a global hybrid. Matthias Ernzerhof joined my group as a postdoc in 1998, and together we built and tested a PBE-based hybrid with the one-quarter exact-exchange fraction that he, Burke, and Perdew had argued for on adiabatic-connection and perturbative grounds.\cite{ref020,ref021} Adamo and Barone published the same functional independently, and their acronym PBE0 is the one in general use.\cite{ref010,ref022}

Global hybrids were already being applied to solids; Muscat, Wander, and Harrison had reported B3LYP band gaps for a set of crystalline solids in 2001.\cite{ref023} An early demonstration of a different kind came in 2002, with Kudin and Richard Martin, when our global PBE hybrid calculation, with 25\% exact exchange, obtained an antiferromagnetic insulating state for UO\textsubscript{2}, whereas LSDA and PBE predicted a ferromagnetic metal.\cite{ref024} This result showed that our periodic Gaussian implementation could bring hybrid DFT to a difficult f-electron solid. It also made band gaps and electronic structure a concrete motivation for what followed. In my recollection, the success sharpened the question posed by the linear-scaling work: could we retain these benefits while making hybrid calculations practical for a much wider range of periodic systems?

\subsection{Two lineages, and an inversion}\label{sec:inversion}

HSE sits at the junction of two conceptual lines. The first is the hybrid functional itself, introduced by Axel Becke in 1993,\cite{ref025} in which a fraction of exact exchange is mixed with a semilocal approximation. Becke's construction is the reason exact exchange entered practical density functional theory, and every functional discussed here that contains a Fock term is downstream of it. Becke died on 23 October 2025;\cite{ref026} the question of how much exact exchange to admit, and on what grounds, is his question before it is anyone else's.

The second line is the range separation of the Coulomb interaction, which I learned from Andreas Savin.\cite{ref010,ref027} Two features of his approach mattered for what followed: the partition itself and the observation that an error-function kernel yields analytic Gaussian two-electron integrals requiring only a trivial modification of the standard formulas. The latter made the construction implementable in an existing code.

Savin's prescription, however, pointed the other way for solids. For molecules, one keeps long-range Fock exchange, which repairs the asymptotic behavior of the potential in the density tail, and assigns the short range to DFT; this was the basis of the long-range-corrected functionals then being developed,\cite{ref028} and later of the CAM and LC families.\cite{ref029,ref030} Solids have no density tails. And Ernzerhof knew from his work with Perdew on the exact exchange hole that the dominant error in the PBE exchange hole is short-ranged.\cite{ref031}

So the prescription was inverted: retain Fock exchange at short range, screen it out at long range, and let semilocal exchange take over. This screening applies to the kernel in the explicit exact-exchange term; the classical Hartree interaction retains the full Coulomb kernel, \(1/r\), via FMM.\cite{ref008} The justification concerns the explicit long-range exchange term specifically. In the uniform electron gas, the long-range part of exact exchange is largely canceled by long-range correlation of the kind resummed in the random-phase approximation.\cite{ref032} A practical functional that includes exact long-range exchange does not include the corresponding long-range correlation, so screening the explicit Fock term is more defensible than retaining it alone. This is a physical motivation for the model, not an exact cancellation built into the functional: HSE screens the Fock term and retains full PBE correlation, and no claim is made that it contains a matching many-body long-range correlation contribution. The argument is also conditional on the screening it assumes. It is strongest for metals and well-screened semiconductors. In weakly screened, wide-gap insulators, the cancellation is incomplete, and suppressing the explicit long-range Fock contribution is a less suitable approximation. That limitation, and the screened-exchange analysis behind it, are taken up in Secs.~\ref{sec:limitations} and~\ref{sec:beyond}. The same motivation does not transfer to properties that depend on the asymptotic potential --- Rydberg and long-range charge-transfer excitations, and response properties of extended molecules --- for which full long-range exact exchange is important; screened hybrids have also been validated for molecular energetics.\cite{heyd2004molecular,ref033} This asymptotic limitation does not imply poor performance for every molecular response property: HSE often retains its parent global hybrid\textquotesingle s performance even for properties sensitive to long-range exchange.\cite{audit357}

\begin{figure}[tb]
\centering
\includegraphics[width=\columnwidth]{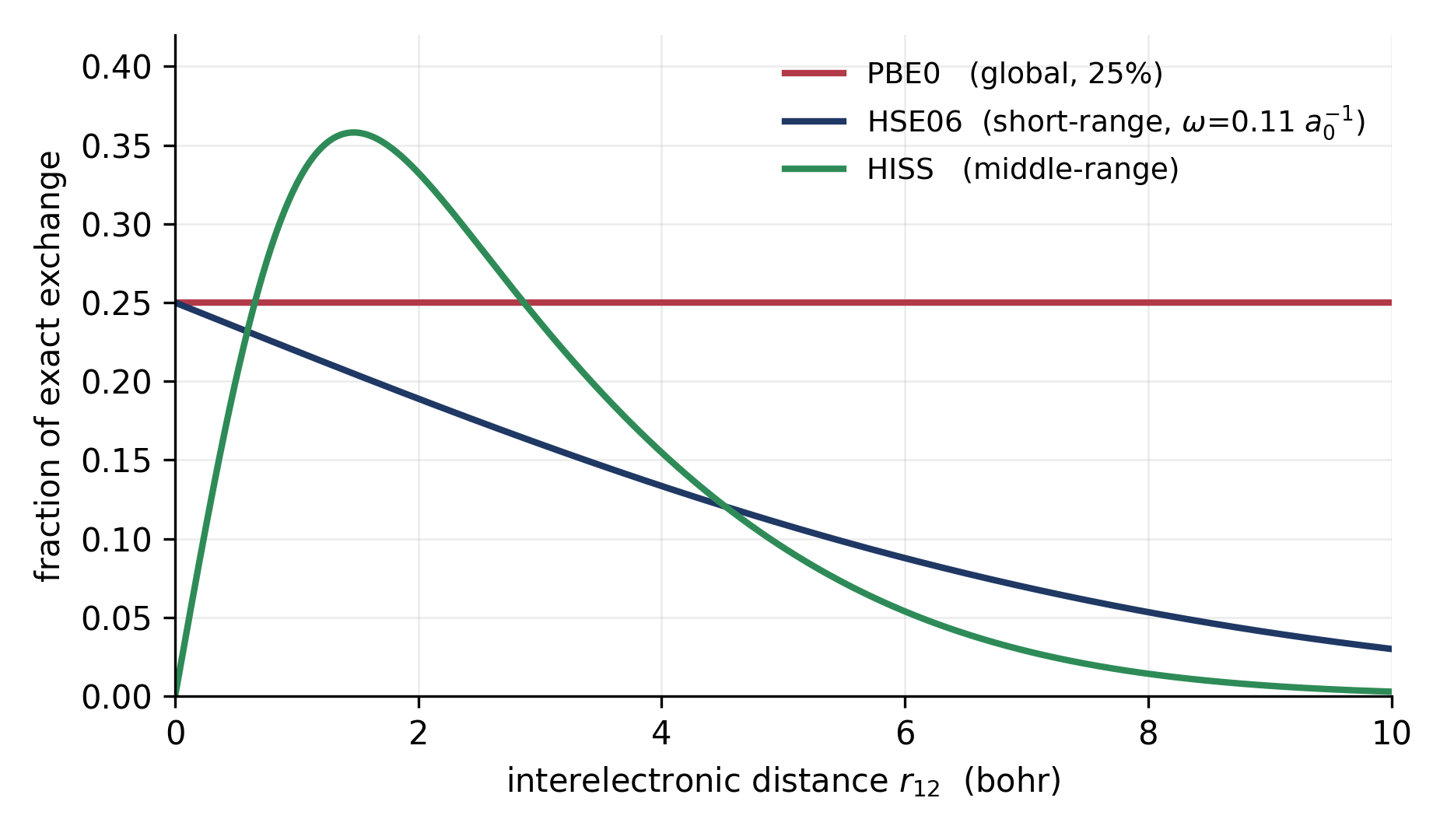}
\caption{Distance-dependent weight of exact exchange for a global hybrid (PBE0, uniform fraction 0.25), a short-range screened hybrid (HSE06, \(\omega\) = 0.11 bohr\(^{-1}\)), and a middle-range screened hybrid (HISS, \(\omega_{\mathrm{SR}}\) = 0.84, \(\omega_{\mathrm{LR}}\) = 0.20 bohr\(^{-1}\), \(c_{\mathrm{MR}}\) = 0.60). The curves show how the distance-dependent weight of exact exchange differs among the three constructions; both the range dependence and the overall weighting differ, and the figure does not separate their individual effects. Computed from the definitions in Refs.~\onlinecite{ref008,ref020,ref033,ref035,ref036}.}
\label{fig:exchange}
\end{figure}

The resulting exchange-correlation energy partitions the Coulomb operator with an error function, \(1/r = \operatorname{erfc}(\omega r)/r + \operatorname{erf}(\omega r)/r\), and mixes exact exchange into the short-range component only:

\begin{equation*}
\begin{aligned}
E_{\mathrm{xc}}^{\mathrm{HSE}} ={}&
 a E_{\mathrm{x}}^{\mathrm{HF,SR}}(\omega)
 +(1-a)E_{\mathrm{x}}^{\mathrm{PBE,SR}}(\omega)\\
 &+E_{\mathrm{x}}^{\mathrm{PBE,LR}}(\omega)
 +E_{\mathrm{c}}^{\mathrm{PBE}},
\end{aligned}
\end{equation*}

where SR and LR denote the short- and long-range components of the partitioned operator, and \(a = 1/4\) with \(\omega\) = 0.11 bohr\(^{-1}\) in HSE06.\cite{ref008,ref033} Figure~\ref{fig:exchange} places this construction against the global hybrid it descends from and the middle-range variant discussed in Sec.~\ref{sec:structures}.

\subsection{The 2003 paper and HSE06}\label{sec:hse06}

Jochen (JJ) Heyd carried out the work; the paper was submitted on 2 December 2002 and published in 2003.\cite{ref008} Tractability was the prime design target, but improved band gaps were already on our agenda after the UO\textsubscript{2} result. Whether removing long-range exact exchange would preserve the benefits seen with the global hybrid remained to be tested. The original paper established the screened construction and demonstrated its computational advantages, including in periodic nanotubes; extensive crystalline benchmarks followed in 2004 and 2005.\cite{ref008,ref037,ref003}

In 2006, Oleg Vydrov and Artur Izmaylov identified an inconsistency in the original implementation.\cite{ref010,ref033} The erratum documented the screening parameters actually used in HSE03: \(\omega_{\mathrm{HF}}=0.15/\sqrt{2}\approx0.106\) bohr\(^{-1}\) in Fock exchange and \(\omega_{\mathrm{PBE}}=0.15\,2^{1/3}\approx0.189\) bohr\(^{-1}\) in PBE exchange.\cite{ref008} Krukau, Vydrov, Izmaylov, and I then examined the screening dependence systematically and recommended the common value of 0.11 bohr\(^{-1}\) that defines HSE06. The correction reduced the mean absolute error in G3/99 formation enthalpies from 6.57 to 4.84 kcal mol\(^{-1}\), while the gap error for 13 semiconductors and insulators changed only from 0.19 to 0.21 eV. Good lattice-constant accuracy was also retained.\cite{ref033}

This preserved gap accuracy was associated with keeping the Fock screening parameter close to its original value while correcting the PBE parameter. Varying the common parameter did affect the gaps: their mean absolute error rose to 0.32 and 0.46 eV at \(\omega=0.15\) and 0.20 bohr\(^{-1}\), respectively. The choice of 0.11 bohr\(^{-1}\) balanced accuracy and computational cost; calculations on solids at smaller values had proved impractically slow in that study.\cite{ref033} The broader aim was a single functional useful for molecular chemistry and solids, exploiting the considerable freedom to vary short- and long-range exchange fractions while retaining molecular accuracy.\cite{audit278} That aim later motivated HISS's three-range construction, whose parameters were fitted to molecular atomization energies and barrier heights together with solid-state band gaps.\cite{ref035,ref083,ref036} The analytic GGA exchange-hole models developed in 2008 facilitated the evaluation of range-separated hybrids.\cite{audit322}

Validation included efficient hybrid calculations in solids,\cite{ref037} a systematic band gap and lattice parameter benchmark,\cite{ref003} the standardized parameter set,\cite{ref033} spin--orbit-corrected gaps,\cite{ref038} comparison against diffusion Monte Carlo for silicon phases and defects,\cite{ref039} solid-state gaps in a periodic Gaussian code,\cite{ref040} and an overview.\cite{ref034}

\section{What HSE Made Possible}\label{sec:impact}

By the time the VASP HSE paper appeared in April 2006, our group and collaborators had established the functional construction, its periodic Gaussian implementation, and systematic tests of its accuracy and computational cost. The 2004 solid-state study assessed 21 metals, semiconductors, and insulators and reported semiconductor gap errors averaging 0.2 eV; the 2005 benchmark covered 40 semiconductors and obtained a mean absolute gap error below 0.3 eV.\cite{ref037,ref003} Molecular validation and early nanotube and actinide applications had already demonstrated the breadth of the approach.\cite{heyd2004molecular,ref060,barone2005metallic,barone2006workfunction,ref046} Among the early applications were systematic studies of narrow nanotube electronic structure in 2004.\cite{audit243} Table~\ref{tab:history} summarizes these early developments. These studies established a working computational method and a body of materials results on which subsequent implementations could build.

Paier and co-workers reported the implementation of HSE in VASP's plane-wave and projector-augmented-wave framework in 2006,\cite{ref044} building on their earlier implementation of full-range exact exchange and molecular tests with PBE0.\cite{paier2005pbe0} We also helped debug the VASP implementation: the Vienna team's erratum records the use of Heyd's semilocal HSE routines and comparisons with our reference exchange energies, which exposed a unit-conversion error.\cite{ref044} The reciprocal-space implementation and the Vienna assessments made an important contribution of their own, helping bring an already developed and tested functional into widespread materials practice.\cite{ref044,ref045} Joachim Paier was then a graduate student in Vienna and later a postdoc in my group. In my view, availability in VASP greatly broadened HSE's reach among researchers working with plane waves; without it, adoption in that community might have been slower.

The collaboration with Richard Martin and colleagues at Los Alamos was a source of scientific questions and applications. Much of the impetus to test HSE on difficult metal oxides came from that collaboration, where the goals included band structures, densities of states, and the nature of metal--oxygen bonding. Ionut Prodan's early Gaussian calculations on uranium and plutonium oxides examined structural, electronic, and magnetic properties together;\cite{ref046} related work extended the assessment to cerium oxides.\cite{ref047} The subsequent study across the actinide dioxide series identified enhanced 5f--O 2p mixing in its intermediate members, associated with the proximity of the orbital energies.\cite{ref046} HSE thus provided a physical account of trends in covalency and improved insulating gaps; this broader program was reviewed by Wen, Martin, Henderson, and me.\cite{ref048}

The experimental comparisons went beyond gap values. Roy and co-workers found that the weak dispersion of uranium 5f-derived bands measured by angle-resolved photoemission in UO\textsubscript{2} agreed with HSE06.\cite{ref049} Later work led by Xiao-Dong Wen extended these studies to spin--orbit coupling and mixed-valence U\textsubscript{3}O\textsubscript{8}.\cite{wen2012soc,wen2013u3o8} Oxygen K-edge absorption measurements on U\textsubscript{3}O\textsubscript{8}, interpreted using the earlier HSE electronic structure, identified oxygen 2p mixing with uranium 5f and 6d states.\cite{wen2013u3o8,wen2014xas} This interaction between theory and spectroscopy made HSE useful for interpreting bonding in f-element oxides.

The wider applications require an important qualification. Host-gap accuracy is necessary for much of what follows, but it is not sufficient: defect levels and band offsets also depend on the description of localized states and on alignment procedures, and two methods reproducing the same host gap can still place a defect level differently.\cite{ref042,ref043}

\begin{table*}[t]
\squeezetable
\caption{Selected milestones from the Scuseria group and collaborators before publication of the VASP HSE implementation in April 2006.\textsuperscript{*}} 
\label{tab:history}
\fontsize{9}{11}\selectfont
\setlength{\tabcolsep}{3pt}
\renewcommand{\arraystretch}{1.10}
\begin{tabularx}{\textwidth}{@{}L{.10\textwidth}@{\hspace{6pt}}L{.24\textwidth}@{\hspace{6pt}}Y@{\hspace{6pt}}C{.065\textwidth}@{}}
\toprule
\textbf{Date} & \textbf{Study} & \textbf{Contribution} & \textbf{Refs.} \\
\midrule
Dec. 2002 & Kudin, Scuseria, and Martin & First periodic hybrid-DFT application to an f-element solid: the global PBE hybrid yields an antiferromagnetic insulating state for UO\textsubscript{2}. & \onlinecite{ref024} \\
\addlinespace[3pt]
May 2003 & Heyd, Scuseria, and Ernzerhof & Introduced HSE and its original periodic Gaussian implementation, with molecular tests and computational demonstrations for semiconducting and metallic nanotubes. & \onlinecite{ref008} \\
\addlinespace[3pt]
Apr. 2004 & Heyd and Scuseria & Extended the original molecular tests to a broad assessment of formation enthalpies, geometries, and vibrational frequencies. & \onlinecite{heyd2004molecular} \\
\addlinespace[3pt]
July 2004 & Heyd and Scuseria & First report of accurate HSE band gaps for a benchmark set of crystalline semiconductors (mean absolute error 0.2 eV), with efficient integral screening and lattice-constant/bulk-modulus tests across 21 solids. & \onlinecite{ref037} \\
\addlinespace[3pt]
Apr. 2005 & Uddin, Peralta, and Scuseria & Applied HSE in periodic Gaussian calculations to bulk and oxygen-deficient PtO, predicting a semiconducting state for bulk PtO where LSDA, PBE, and TPSS gave a metal. & \onlinecite{audit249} \\
\addlinespace[3pt]
July--Aug. 2005 & Barone, Peralta, and co-workers & Extended HSE to quantitative optical-transition calculations for semiconducting and metallic nanotubes. & \onlinecite{ref060,barone2005metallic} \\
\addlinespace[3pt]
Oct. 2005 & Heyd, Peralta, Scuseria, and Martin & Established a 40-semiconductor HSE benchmark in Gaussian, demonstrating a mean absolute gap error below 0.3 eV together with lattice-constant accuracy. & \onlinecite{ref003} \\
\addlinespace[3pt]
Dec. 2005 & Nakai, Heyd, and Scuseria & Early HSE03 validation for transition-metal oxides: gaps, cohesion, structures, and efficient exchange truncation in periodic Gaussian calculations on anatase and rutile TiO\textsubscript{2}. & \onlinecite{nakai2006tio2} \\
\addlinespace[3pt]
Jan. 2006 & Prodan, Scuseria, and Martin & Established HSE results for uranium and plutonium oxides in Gaussian, assessing structural, electronic, and magnetic properties together. & \onlinecite{ref046} \\
\addlinespace[3pt]
Jan. 2006 & Barone, Peralta, Uddin, and Scuseria & Extended HSE to work-function predictions for pristine and doped carbon nanotubes. & \onlinecite{barone2006workfunction} \\
\bottomrule
\end{tabularx}
\par\vspace{3pt}
\begin{minipage}{\textwidth}\footnotesize\raggedright
\textsuperscript{*}The HSE applications listed here were carried out at Rice before publication of the VASP HSE implementation in April 2006.\cite{ref044} Dates refer to the first online publication. The 2002 UO\textsubscript{2} calculation is a global-hybrid precursor; the early HSE results predate the HSE06 parametrization. \par\vspace{2pt}
\end{minipage}
\end{table*}

Table~\ref{tab:i} lists representative application domains with the protocol actually used in each cited study, since standard HSE06, system-tuned exchange fractions, and global hybrids are not interchangeable.

\begin{table*}[t]
\squeezetable
\caption {Representative applications of HSE and related screened hybrid methods across the research community, with the protocol used in each cited work.\textsuperscript{*}} 
\label{tab:i}
\fontsize{9}{11}\selectfont
\setlength{\tabcolsep}{3pt}
\renewcommand{\arraystretch}{1.12}
\begin{tabularx}{\textwidth}{@{}L{.15\textwidth}@{\hspace{6pt}}Y@{\hspace{6pt}}L{.22\textwidth}@{\hspace{6pt}}C{.075\textwidth}@{}}
\toprule
\textbf{Domain} & \textbf{What the cited work reports} & \textbf{Protocol} & \textbf{Refs.} \\
\midrule
Defect levels & Semilocal and global-hybrid defect levels agree closely once aligned to a common reference; band-edge placement is the limiting factor & PBE vs PBE0; tuned PBEh & \onlinecite{ref042,ref043,ref050} \\
\addlinespace[3pt]
Solid-state qubits & High-throughput screening of candidate spin-photon interfaces in silicon & Single-shot HSE on PBE states; self-consistent HSE06 refinement & \onlinecite{ref051} \\
\addlinespace[3pt]
Polaron localization & Stabilization of small polarons and the associated lattice distortion, which self-interaction error suppresses & HSE06 & \onlinecite{ref052} \\
\addlinespace[3pt]
Halide perovskites & HSE with spin--orbit coupling reveals deep levels and nonradiative centers, prompting reassessment of the earlier defect-tolerance account obtained without SOC & HSE06 + SOC & \onlinecite{ref053} \\
\addlinespace[3pt]
Wide-gap nitrides & Identification of carbon-related centers as a source of yellow luminescence in GaN & Tuned HSE, a = 0.31 & \onlinecite{ref054} \\
\addlinespace[3pt]
Band offsets & Offsets in nitride heterostructures; oxide interfaces treated with system-tuned fractions & HSE06;\cite{ref055} tuned\cite{ref056} & \onlinecite{ref055,ref056} \\
\addlinespace[3pt]
Photocatalysis & Band-edge alignment to redox levels, obtained within a protocol combining hybrid and embedded-cluster calculations with experimental gaps & HSE06 within a mixed protocol & \onlinecite{ref057} \\
\addlinespace[3pt]
Battery cathodes & Redox potentials and formation energies without species-specific U; a related study calibrates system-specific fractions & HSE06;\cite{ref058} tuned\cite{ref059} & \onlinecite{ref058,ref059} \\
\addlinespace[3pt]
Actinide and lanthanide oxides & Insulating gaps and improved density-of-states placement; the 2002 UO\textsubscript{2} result used a global hybrid and is a historical predecessor & HSE; PBE0 in Ref.~\onlinecite{ref024} & \onlinecite{ref024,ref046,ref047,ref048,ref049} \\
\addlinespace[3pt]
Low-dimensional systems & Electronic gaps and indirect-to-direct transitions in nanotubes, graphene nanoribbons, and multilayer MoS\textsubscript{2}; optical excitation energies require excitonic treatment & HSE-family calculations & \onlinecite{ref060,ref061,ref062,ref063} \\
\bottomrule
\end{tabularx}
\par\vspace{3pt}
\begin{minipage}{\textwidth}\footnotesize\raggedright
\textsuperscript{*} HSE06 denotes the standard parameter set; ``tuned'' denotes a system-specific exchange fraction. The recurring pattern is that semilocal functionals provide useful structures and trends but can be unreliable when predictions depend sensitively on band-edge placement or on the localization of a defect, polaron, or d or f electron. \par\vspace{2pt}
\end{minipage}
\end{table*}

Point-defect thermodynamics is the clearest case of a field reorganized around these calculations: the standard review treats hybrid-functional results as a methodological reference.\cite{ref050}

It is worth recording what else 2003 contained, because it bears on the second part. The same year produced a practical local-hybrid construction with a position-dependent mixing fraction,\cite{ref064} a form anticipated by earlier work.\cite{ref065} It expressed the aim of adapting exact exchange to the electronic environment. A subsequent extension in 2008, developed with Perdew and Savin, made the range-separation parameter itself a local function of position,\cite{ref066} an early step toward the material-dependent screening discussed in Sec.~\ref{sec:developments}. A complementary 2008 construction made the fraction of screened exchange position dependent; its initial tests were molecular.\cite{audit333} TPSS, the nonempirical meta-GGA written with Jianmin Tao, John Perdew, and Viktor Staroverov, also appeared in 2003, building on the earlier PKZB functional.\cite{ref067,ref068} My collaboration with John Perdew, which began there and includes PBEsol,\cite{ref069} has run in parallel with the hybrid work for two decades. TPSS is a conceptual ancestor of SCAN and r\textsuperscript{2}SCAN, and therefore of the meta-GGAs that now occupy the same territory. HSE and TPSS both appeared in 2003, advancing two already developing approaches that now meet in band-gap prediction.

\section{What HSE Did Not Do}\label{sec:limitations}

The most instructive criticism arrived early and from a friendly quarter. At conferences during the actinide period, Hardy Gross would point out that a broken-symmetry single-determinant hybrid does not model the short-range entanglement from which the gap of a Mott insulator arises, so agreement in those systems does not establish that the underlying physics has been captured.\cite{ref010} He was correct, and the point concerns what this class of approximations demonstrates rather than what density functional theory can, in principle, achieve. A useful number for UO\textsubscript{2} is not an explanation of UO\textsubscript{2}, and system-specific claims about particular Mott systems require system-specific evidence.

The technical limitations are definite. The exact-exchange fraction is fixed at one quarter. Model arguments relate an appropriate fraction to inverse dielectric screening,\cite{ref070,ref071} and while that rationale is model-dependent --- the range parameter also matters, and no single universal fraction follows --- the fixed profile does leave HSE substantially underestimating wide-gap insulators, as Sec.~\ref {sec:beyond} documents. The underlying reason is the one given in Sec.~\ref{sec:inversion}. Weaker electronic screening leaves a larger residual long-range screened-exchange contribution than HSE's fixed profile represents. This limitation belongs to the standard parameterization rather than to screened hybrids as a class: material-dependent tuning of the short-range exchange fraction can improve wide-gap predictions while retaining good semiconductor gaps.\cite{ref041,ref070} The fixed-profile benchmark comparisons in Sec.~\ref{sec:beyond}, however, do not separately identify the effects of exchange fraction and range. The screening length is likewise fixed, so the functional cannot adapt to materials whose screening differs qualitatively, reduced-dimensional systems being the obvious modern case. And a screened hybrid remains more expensive than a semilocal functional, by an amount that depends far more on implementation than is generally appreciated (Sec.~\ref{sec:structures}).

A caution about benchmarking applies to my own work as much as to anyone else's. In a study of four hybrids on 41 compounds, of which 27 converged for all four functionals, we found that different error measures rank the same functionals differently, and a Wilcoxon signed-rank test on those 27 compounds found no significant difference among HSE, B3PW91, and B3LYP.\cite{ref041} That is a statement about one set of that size, not a universal rule about any particular error threshold. We also found the experimental reference data for correlated oxides to be uncertain by more than the differences under discussion: reported gaps for CoO span 2.1 to 5.43 eV, with a standard deviation over six literature values of 1.22 eV. Every comparison in the second part, including those favorable to screened hybrids, should be read against that background.

\section*{\texorpdfstring{\makebox[\columnwidth]{The Contenders}}{The Contenders}}

\section{The Challenge}\label{sec:challenge}

Two recent meta-GGAs have reopened a question that seemed settled. LAK, developed by Lebeda, Aschebrock, and Kümmel, exploits an unused freedom in the second-order gradient expansion, redistributing weight between the density gradient and the kinetic energy density, and reports a mean absolute error of 0.18 eV for fifteen sp semiconductors against 0.17 eV for HSE06 in the same calculation, with no Fock exchange.\cite{ref072} Wang, Shahi, Perdew, and Ruzsinszky's bn-r\textsuperscript{2}SCAN fits a more flexible interpolation function to band gaps and lattice constants together, and in doing so analyzes how those two properties compete within the meta-GGA form.\cite{ref073} Both are fine pieces of work, and the obvious question is how far these methods reproduce the capabilities of screened hybrids.

This Perspective takes that question narrowly. HSE offered a combination of gaps, structures, energetics, and affordability, and the issue is which parts of it the newer methods reproduce, and on what evidence. The comparisons below are restricted to methods benchmarked on sets of solids, which excludes a large body of literature developed and assessed for molecules; this exclusion reflects what can be compared on common ground, not a judgment of importance.

Three things must be controlled before any such comparison means anything. The target quantity: a bound excitonic excitation lies below the corresponding quasiparticle transition by its binding energy. Direct or indirect character, temperature, phonon assistance, zero-point renormalization --- roughly 0.4 eV for diamond --- and spin--orbit coupling, which lowers the computed gap of PbTe by more than half an electron volt, all require a stated convention.\cite{ref074} The reference data: for correlated oxides, the experimental spread exceeds the differences under discussion, reported gaps for CoO spanning 2.1 to 5.43 eV.\cite{ref041} And the geometry: most of the benchmarks below use fixed experimental structures, which isolate the electronic approximation, while a few use each method's own relaxed cell,\cite{ref036} which is the more demanding test and the one relevant to prediction.

\section{Semiconductors: The Contention Holds}\label{sec:semiconductors}

\begin{table*}[t]
\squeezetable
\caption{Band gap mean absolute errors for sp semiconductors. SCBG15 comprises fifteen materials with gaps between 0.5 and 4.0 eV; SC40 group 1 comprises twenty semiconductors with low-temperature references, for which not every functional converged. The SCBG15 entries from Ref.~\onlinecite{ref072} are mutually controlled scalar-relativistic calculations without spin--orbit coupling. The two SC40 entries from Ref.~\onlinecite{ref036} are also mutually controlled.}
\label{tab:ii}
\fontsize{9}{11}\selectfont
\setlength{\tabcolsep}{3pt}
\renewcommand{\arraystretch}{1.12}
\begin{tabularx}{\textwidth}{@{}L{.14\textwidth}@{\hspace{6pt}}C{.28\textwidth}@{\hspace{6pt}}C{.045\textwidth}@{\hspace{6pt}}C{.20\textwidth}@{\hspace{6pt}}C{.18\textwidth}@{\hspace{6pt}}C{.065\textwidth}@{}}
\toprule
\textbf{Method} & \textbf{Set} & \textbf{N} & \textbf{Geometry} & \textbf{MAE (eV)} & \textbf{Ref.} \\
\midrule
HSE06 & SCBG15 & 15 & experimental & 0.17 & \onlinecite{ref072} \\
LAK & SCBG15 & 15 & experimental & 0.18 & \onlinecite{ref072} \\
bn-r\textsuperscript{2}SCAN & SCBG15 (its fitting set) & 15 & experimental & 0.28 & \onlinecite{ref073} \\
r\textsuperscript{2}SCAN & SCBG15 & 15 & experimental & 0.52 & \onlinecite{ref073} \\
SCAN & SCBG15 & 15 & experimental & 0.59 & \onlinecite{ref072} \\
PBE & SCBG15 & 15 & experimental & 0.91 & \onlinecite{ref072} \\
\midrule
HSE & SC40 group 1 & 20 & experimental & 0.18 & \onlinecite{ref041} \\
HSE & SC40 group 1 & 20 & relaxed & 0.21 & \onlinecite{ref036} \\
HISS & SC40 group 1 & 20 & relaxed & 0.39 & \onlinecite{ref036} \\
PBE0 & SC40 group 1 & 17 & experimental & 0.55 / 0.58 & \onlinecite{ref041} \\
\bottomrule
\end{tabularx}
\par\vspace{3pt}
\begin{minipage}{\textwidth}\footnotesize
The PBE0 entry covers 17 of the 20 group-1 compounds, \(\beta\)-SiC, BP, and InN being unavailable. Two statistics are given: 0.55 eV as published in Ref.~\onlinecite{ref041}, and 0.58 eV as recomputed from the rounded values in its supplementary table.
\end{minipage}
\end{table*}

For selected semiconductors, LAK supports the contention that a semilocal functional can achieve HSE06-level band-gap accuracy. On SCBG15 (Table~\ref{tab:ii}), LAK and HSE06 have similar reported mean absolute errors, 0.18 and 0.17 eV, respectively.\cite{ref072} The 0.01 eV difference neither supports a meaningful ranking nor establishes statistical equivalence. HSE also gives an MAE of 0.18 eV on a different set in a different code and basis.\cite{ref041} The bn-r\textsuperscript{2}SCAN entry is in-sample, SCBG15 being its fitting target.\cite{ref073}

One piece of context is usually missing from this discussion. A semilocal method reached screened-hybrid gap accuracy more than a decade ago: the modified Becke--Johnson potential, with parameters fitted to experimental gaps,\cite{ref075} was among the most accurate of 21 functionals tested on 472 solids, alongside HSE06.\cite{ref004,ref076} What mBJ cannot supply is total energies, forces, or stresses, since it is not the functional derivative of an energy. SCAN, r\textsuperscript{2}SCAN, and TASK are meta-GGA energy functionals.\cite{ref077,ref078,ref079} LAK's achievement is to reach HSE06-level accuracy for these semiconductors within that variational framework.\cite{ref072} The mechanism is the same generalized Kohn--Sham effect that hybrids exploit, achieved via the kinetic energy density rather than nonlocal exchange.\cite{ref007}

\section{Beyond Semiconductors}\label{sec:beyond}

Outside that window, the picture changes, and this is where the comparison with HSE becomes informative rather than merely close.

\begin{table*}[t]
\squeezetable
\caption{Band gaps on chemically broader sets. The 24-material set spans 0.3--6.4 eV and includes closed-shell \(d^0\) and \(d^{10}\) oxides together with seven correlated transition-metal and rare-earth oxides. The three-, two-, and one-dimensional sets are subsets of a 100-material study referenced to GW rather than to experiment.}
\label{tab:iii}
\fontsize{9}{11}\selectfont
\setlength{\tabcolsep}{3pt}
\renewcommand{\arraystretch}{1.12}
\begin{tabularx}{\textwidth}{@{}L{.10\textwidth}@{\hspace{6pt}}C{.235\textwidth}@{\hspace{6pt}}C{.04\textwidth}@{\hspace{6pt}}C{.10\textwidth}@{\hspace{6pt}}C{.14\textwidth}@{\hspace{6pt}}C{.125\textwidth}@{\hspace{6pt}}C{.085\textwidth}@{\hspace{6pt}}C{.055\textwidth}@{}}
\toprule
\textbf{Method} & \textbf{Set} & \textbf{N} & \textbf{Reference} & \textbf{Geometry} & \textbf{MAE (eV)} & \textbf{ME (eV)} & \textbf{Ref.} \\
\midrule
LAK & 24 materials & 24 & experiment & experimental & 0.80 & \ensuremath{-}0.77 & \onlinecite{ref073} \\
bn-r\textsuperscript{2}SCAN & 24 materials & 24 & experiment & experimental & 0.85 & \ensuremath{-}0.82 & \onlinecite{ref073} \\
r\textsuperscript{2}SCAN & 24 materials & 24 & experiment & experimental & 1.03 & \ensuremath{-}1.03 & \onlinecite{ref073} \\
HSE06 & 24 materials & 24 & experiment & experimental & not reported & --- & \onlinecite{ref073} \\
\midrule
HSE06 & 33 bulk (3D) & 33 & GW & not stated & 1.13 & \ensuremath{-}1.11 & \onlinecite{ref080} \\
LAK & 33 bulk (3D) & 33 & GW & not stated & 1.73 & \ensuremath{-}1.71 & \onlinecite{ref080} \\
HSE06 & 33 two-dimensional & 33 & GW & not stated & 0.67 & --- & \onlinecite{ref080} \\
LAK & 33 two-dimensional & 33 & GW & not stated & 0.95 & --- & \onlinecite{ref080} \\
HSE06 & 34 one-dimensional & 34 & GW & not stated & 2.03 & --- & \onlinecite{ref080} \\
LAK & 34 one-dimensional & 34 & GW & not stated & 2.47 & --- & \onlinecite{ref080} \\
\midrule
HSE & SC40 common, \ensuremath{<}8 eV & 25 & experiment & experimental & 0.26 & --- & \onlinecite{ref041} \\
\bottomrule
\end{tabularx}
\par\vspace{3pt}
\begin{minipage}{\textwidth}\footnotesize
The 25-material row is the 27-compound set on which all four hybrids of Ref.~\onlinecite{ref041} converged, with NaCl and LiCl removed as the compounds above roughly 8 eV. The geometry convention is not specified in the available main text of Ref.~\onlinecite{ref080}; these rows are therefore kept separate from benchmarks with stated structural conventions.
\end{minipage}
\end{table*}

LAK's error on the 24-material set (Table~\ref{tab:iii}) is four times its SCBG15 value, with the magnitude of the mean error close to the mean absolute error, so underestimation dominates. That block contains no matched HSE06 entry and cannot, by itself, establish a hybrid advantage; what it shows is that all three meta-GGAs underestimate substantially once oxides enter, with LAK marginally the best of them. The matched comparisons yield larger mean absolute errors for LAK than for HSE06 across all three dimensionalities, relative to the GW references.\cite{ref080} Both functionals are compared with the same GW reference data within each set, establishing their relative agreement with those calculations; the corresponding ranking against experiment requires separate validation.

The most direct evidence against experiment comes from Riemelmoser, Xu, and Pasquarello's 39-material benchmark spanning PbTe to solid neon, evaluated at experimental geometries with that study's spin--orbit and zero-point corrections.\cite{ref081} By mean absolute relative error, HSE gives 16\% and LAK 33\%; LAK is the best of the semilocal functionals, ahead of r\textsuperscript{2}SCAN at 46\% and PBE at 63\%, and remains a factor of two behind the screened hybrid.

\begin{table*}[t]
\squeezetable
\caption{Band gap accuracy over the 39-material benchmark of Ref.~\onlinecite{ref081}, at experimental geometries with that study's stated spin--orbit and zero-point corrections. MARE is the mean absolute relative error. ``Dielectric input'' identifies how the exchange fraction was obtained.}
\label{tab:iv}
\fontsize{9}{11}\selectfont
\setlength{\tabcolsep}{3pt}
\renewcommand{\arraystretch}{1.12}
\begin{tabularx}{\textwidth}{@{}L{.19\textwidth}@{\hspace{6pt}}C{.27\textwidth}@{\hspace{6pt}}C{.105\textwidth}@{\hspace{6pt}}C{.105\textwidth}@{\hspace{6pt}}Z@{}}
\toprule
\textbf{Method} & \textbf{Dielectric input} & \textbf{MARE (\%)} & \textbf{MAE (eV)} & \textbf{Character} \\
\midrule
PBE & --- & 63 & --- & semilocal \\
r\textsuperscript{2}SCAN & --- & 46 & --- & semilocal meta-GGA \\
LAK & --- & 33 & --- & semilocal meta-GGA \\
PBE0 & fixed \(a = 1/4\) & 44 & 0.79 & global hybrid \\
r\textsuperscript{2}SCAN0 & fixed \(a = 1/4\) & 64 & 0.84 & global hybrid \\
HSE & fixed \(a = 1/4\) & 16 & 0.94 & screened hybrid \\
DD-PBEH & experimental & 27 & 0.37 & global dielectric-dependent \\
DD-r\textsuperscript{2}SCANH & experimental & 7 & 0.19 & global dielectric-dependent \\
DD-PBEH & calculated (PBE) & 35 & 0.50 & global dielectric-dependent \\
DD-r\textsuperscript{2}SCANH & calculated (r\textsuperscript{2}SCAN) & 10 & 0.18 & global dielectric-dependent \\
\bottomrule
\end{tabularx}
\par\vspace{3pt}
\begin{minipage}{\textwidth}\footnotesize
Values follow Table 4 and Supplementary Note 5 of Ref.~\onlinecite{ref081}. Many of the fixed-quarter and calculated-dielectric global-hybrid gaps in that table were obtained by interpolation with respect to the exchange fraction; all narrow-gap semiconductors were calculated explicitly. The PBE entry follows that table; the prose of the same work gives 62\%.
\end{minipage}
\end{table*}

Two features of Table~\ref{tab:iv} bear directly on the first part. The first is that the set reaches solid neon, and its wide-gap end is where the screening rationale of Sec.~\ref{sec:inversion} is least favorable to HSE. In weakly screened insulators, the cancellation of long-range exchange by correlation is incomplete, whereas HSE removes the explicit long-range Fock term entirely. In the macroscopic screened-exchange picture, a three-dimensional insulator with finite electronic dielectric constant \(\varepsilon_\infty\) retains a long-range interaction of order \(1/(\varepsilon_\infty r)\), which the standard HSE profile does not represent.\cite{ref034,ref081} The large absolute errors at that end are consistent with this limitation of standard HSE. Material-dependent tuning can improve wide-gap predictions while retaining good semiconductor gaps, so this limitation does not apply to screened hybrids as a class.\cite{ref041,ref070}

The second is a caution about error measures. On these same 39 materials, HSE has the best relative error among the fixed-fraction hybrids and the worst absolute error, 0.94 eV, compared with 0.79 eV for PBE0 and 0.84 eV for r\textsuperscript{2}SCAN0. The relative measure is carried by the semiconductors and the absolute measure by the wide-gap insulators. Both statements describe the same calculations, and either alone would misrepresent them --- the difficulty documented on a different set in Ref.~\onlinecite{ref041}, where ten error measures ranked four hybrids differently, and a Wilcoxon signed-rank test found no significant difference among three of them.

\section{Structures, Energetics, And Cost}\label{sec:structures}

Gaps are only part of what HSE offered; the remainder is where the two contenders differ most.

\begin{table*}[t]
\squeezetable
\caption{Equilibrium structures. The first two columns report mean absolute errors in lattice constants across different sets; the remainder report mean absolute relative errors on an independent structural set of 20 materials, broken down by class. Values in different columns come from different studies.}
\label{tab:v}
\fontsize{9}{11}\selectfont
\setlength{\tabcolsep}{3pt}
\renewcommand{\arraystretch}{1.12}
\begin{tabularx}{\textwidth}{@{}L{.15\textwidth}@{\hspace{6pt}}C{.12\textwidth}@{\hspace{6pt}}C{.19\textwidth}@{\hspace{6pt}}C{.13\textwidth}@{\hspace{6pt}}C{.13\textwidth}@{\hspace{6pt}}C{.13\textwidth}@{\hspace{6pt}}C{.065\textwidth}@{}}
\toprule
\textbf{Method} & \textbf{LC20 (\AA{})} & \textbf{SC40, 43 params (\AA{})} & \textbf{20 solids (\%)} & \textbf{\shortstack{Insulators\\(\%)}} & \textbf{\shortstack{Metals\\(\%)}} & \textbf{Refs.} \\
\midrule
SCAN & 0.015 & --- & --- & --- & --- & \onlinecite{ref072} \\
HISS & --- & 0.022 & --- & --- & --- & \onlinecite{ref036} \\
PBEsol & --- & 0.024 & --- & --- & --- & \onlinecite{ref036} \\
r\textsuperscript{2}SCAN & 0.026 & --- & 0.67 & 0.30 & 1.53 & \onlinecite{ref073} \\
bn-r\textsuperscript{2}SCAN & 0.026 & --- & 0.84 & 0.53 & 1.57 & \onlinecite{ref073} \\
HSE06 & 0.032 & 0.039 & --- & --- & --- & \onlinecite{ref036,ref072} \\
LAK & 0.054 & --- & 1.58 & 1.03 & 2.84 & \onlinecite{ref072,ref073} \\
PBE & 0.055 & --- & --- & --- & --- & \onlinecite{ref072} \\
\bottomrule
\end{tabularx}
\end{table*}

LAK's gap improvement is accompanied by less accurate lattice constants: its LC20 error is close to PBE's, and on the independent set, it is more than twice r\textsuperscript{2}SCAN's and the worst of the three meta-GGAs for insulators and metals alike. bn-r\textsuperscript{2}SCAN matches r\textsuperscript{2}SCAN's aggregate LC20 accuracy while improving the gaps less. Molecular energetics follow the same pattern of compromise rather than uniform loss: bn-r\textsuperscript{2}SCAN degrades covalent atomization energies, with the W4-11 error rising from 3.8 to 5.6 kcal/mol, while improving barrier heights, dispersion-bound dimers, and water clusters, the WATER27 error falling from 5.07 to 1.04 kcal/mol;\cite{ref073} and a subsequent assessment of LAK reports accurate weak interactions near equilibrium and the best GMTKN55 performance among the pure semilocal-cost functionals tested, without an added dispersion correction.\cite{ref082}

Reference~\onlinecite{ref073} presents gaps and structures as competing objectives within the r\textsuperscript{2}SCAN-type form, and within that form, the analysis is convincing. Whether the competition follows from band-gap prediction as such is another matter. mBJ is the limiting case, buying excellent gaps by abandoning the energy functional altogether. HISS provides an illustration of the hybrid rung: a middle-range screened hybrid whose three parameters were fitted to molecular and solid-state data, yielding 0.022 \AA{} over 43 SC40 lattice parameters and 0.39 eV gaps for group-1 semiconductors, both at its own relaxed geometries.\cite{ref035,ref036,ref083} Its gap accuracy is worse than HSE's on the matched set, and its structural accuracy on that set exceeds PBEsol's; cross-set comparison with Table~\ref{tab:v} is not controlled. The screened-hybrid family is broader than these two members, with M06-SX, N12-SX, and HSE12s populating it at differing levels of empiricism.\cite{ref084,ref085} What the example supports is narrow but relevant: good structures and useful gaps can coexist in a single self-consistent energy functional, and the position of exact exchange in \(r_{12}\) shifts the balance between them.

\begin{table*}[t]
\squeezetable
\caption{Reported CPU time per SCF cycle relative to PBE, with the number of test systems entering each mean. The two blocks are from different codes, basis sets, and test systems, and are not mutually controlled.}
\label{tab:vi}
\fontsize{9}{11}\selectfont
\setlength{\tabcolsep}{3pt}
\renewcommand{\arraystretch}{1.12}
\begin{tabularx}{\textwidth}{@{}L{.14\textwidth}@{\hspace{6pt}}C{.21\textwidth}@{\hspace{6pt}}C{.095\textwidth}@{\hspace{6pt}}Z@{\hspace{6pt}}C{.065\textwidth}@{}}
\toprule
\textbf{Method} & \textbf{Cost rel. to PBE} & \textbf{N} & \textbf{Code/basis} & \textbf{Ref.} \\
\midrule
HISS & 1.91\ensuremath{\times} (1.83\ensuremath{\times}) & 6 (4) & GAUSSIAN, Gaussian orbitals & \onlinecite{ref036} \\
HSE06 & 2.64\ensuremath{\times} & 6 & GAUSSIAN, Gaussian orbitals & \onlinecite{ref036} \\
PBE0 & 35.5\ensuremath{\times} & 4 & GAUSSIAN, Gaussian orbitals & \onlinecite{ref036} \\
\midrule
TASK & 3--4\ensuremath{\times} & n.s. & BAND, Slater orbitals & \onlinecite{ref086} \\
HSE & 50--115\ensuremath{\times} & n.s. & BAND, Slater orbitals & \onlinecite{ref086} \\
\bottomrule
\end{tabularx}
\par\vspace{3pt}
\begin{minipage}{\textwidth}\footnotesize
The main HISS and HSE06 means cover six systems. The parenthetical HISS value is the mean over the four systems used for the matched comparison with PBE0. ``n.s.'' means ``not stated.''
\end{minipage}
\end{table*}

The cost argument for semilocal functionals deserves more scrutiny than it usually receives. Reported penalties for the same functional differ by more than an order of magnitude, and no controlled conclusion can be drawn across the two blocks of Table~\ref{tab:vi}, which differ in code, basis, and systems. What the table does establish is that the penalty for screened exchange is not determined by the functional form alone: in a Gaussian-orbital code, HISS costs less than twice as much as PBE, and HSE06 less than three times as much. Within the first block, the four-system comparison is controlled, and PBE0 costs about 19 times as much as HISS in the same code and systems --- a difference arising from how the Coulomb operator is partitioned and weighted. These timings show that screening can greatly reduce the exchange cost in this implementation; end-to-end cost remains system- and algorithm-dependent, and range separation is not the only route to efficient exact exchange.

\section{Where The Limitation Is Being Addressed}\label{sec:developments}

The fixed exchange fraction identified in Sec.~\ref{sec:limitations} is being addressed directly, and by methods closer to HSE than to the meta-GGAs. Dielectric-dependent hybrids set the fraction from the material's own screening rather than fixing it in advance.\cite{ref070,ref071} The global dielectric-dependent hybrid DD-r\textsuperscript{2}SCANH reaches 7\% mean absolute relative error over the 39-material set with the experimental dielectric constant and 10\% with the constant computed from r\textsuperscript{2}SCAN, against 27\% and 35\% for the PBE-based version and 16\% for HSE.\cite{ref081} Screened range-separated hybrids with nonempirical tuning pursue the same end. An earlier demonstration by Refaely-Abramson and co-workers showed how dielectric screening in a range-separated hybrid captures polarization-induced gap renormalization in molecular crystals.\cite{ref087} Wing and co-workers used Wannier-localization-based optimal tuning for bulk solids, obtaining a mean absolute error of 0.08 eV over 16 semiconductors and insulators against experimental references adjusted for zero-point renormalization, with spin--orbit coupling included where relevant.\cite{ref088} Jana and co-workers' dielectric-dependent construction reports 0.30 eV relative to GW for 33 bulk solids, whereas HSE06 gives 1.13 eV.\cite{ref080} These studies use different material sets and reference conventions, so their aggregate errors do not provide a common ranking. These methods carry costs standard HSE does not: material-specific parameter determination and, in some cases, an additional dielectric calculation.

Two observations follow. The first is that replacing PBE by r\textsuperscript{2}SCAN improves the dielectric-dependent hybrid even when both are given the same experimental screening parameter,\cite{ref081} so progress on the meta-GGA rung improves hybrids as well as standalone semilocal calculations; the two lines cooperate more than they compete. The second is that these constructions pursue, with modern machinery, the aim of the position-dependent mixing and local range-separation work described in Sec.~\ref{sec:impact} --- adapting exact exchange to the electronic environment rather than fixing it once.

A word on the reference standard. GW does not denote a single accuracy: quasiparticle self-consistent GW overestimates gaps with a mean error of +0.55 eV over 154 materials, while quasiparticle self-consistent GW with vertex corrections in W gives a mean error of +0.14 eV and a mean absolute error of 0.31 eV over a different subset of 101 materials.\cite{ref074,ref089} These aggregate statistics are not a matched comparison. On a common 94-material subset, one-shot \(G_0W_0\) from an LDA starting point, using a plasmon-pole approximation, yields an MAE of 0.46 eV, compared with 0.58 eV for mBJ and 0.82 eV for HSE06, while on larger unmatched sets it compares less favorably with mBJ.\cite{ref074} Any functional benchmarked against GW requires the variant, starting point, and subset to be named.

A common benchmark of LAK, bn-r\textsuperscript{2}SCAN, mBJ, HSE06, HISS, and a dielectric-dependent hybrid would help clarify their relative strengths. Consistent numerical treatments, especially for meta-GGA pseudopotentials,\cite{ref090}. Results grouped by material class, with clearly defined reference gaps and both absolute and relative errors, would help distinguish systematic limitations from benchmark-dependent rankings.\cite{ref041,ref074}

In summary, these results demonstrate substantial progress in band-gap prediction within hybrid DFT through material-dependent treatments of exchange and screening.\cite{ref081,ref088,ref080} The global dielectric-dependent and screened range-separated hybrids highlighted here all adapt exact exchange to the material's screening. The question these developments pose is therefore not whether to screen exact exchange in an extended system, but how to choose the screening, building along the road that HSE started.

\section{Outlook}\label{sec:outlook}

LAK is a genuine conceptual advance. It exploits previously unused freedom in the gradient expansion to obtain HSE06-level semiconductor gaps within a variational energy functional,\cite{ref072} with subsequent evidence of good molecular transferability.\cite{ref082} bn-r\textsuperscript{2}SCAN recovers the structural accuracy that LAK gives up, with a smaller, fitted gap improvement.\cite{ref073} What has not been shown is parity with screened hybrids across material classes: on the 39-material benchmark against experiment, LAK trails HSE by a factor of two in relative error,\cite{ref081} and the matched GW comparisons run the same way in three, two, and one dimensions.\cite{ref080} Comparisons against well-characterized experimental gaps for correlated magnetic oxides, and at each method's own relaxed geometry, would clarify how far the gains transfer.

There is a symmetry worth noting. HSE and TPSS both appeared in 2003, advancing two already developing approaches that now meet in band-gap prediction, and the functionals presently competing descend from both. That is less a rivalry between camps than two branches of one program, pursued by overlapping groups of people, reaching comparable semiconductor accuracy by different routes --- one by choosing where to place the exact exchange, the other by shaping the dependence on kinetic energy density. They respond to different constraints and fail in different places, which is why both are worth having.

\section*{\texorpdfstring{\makebox[\columnwidth]{Conclusion}}{Conclusion}}

HSE began with the aim of making exact exchange both practical and physically defensible in extended systems. Its contribution emerged from the combination of a functional construction, a working periodic implementation, and systematic validation for molecules and solids. That sequence, developed at Rice, established a method with demonstrated accuracy and computational advantages that subsequent implementations and applications could build on.\cite{ref008,heyd2004molecular,ref037,ref003} The VASP implementation and independent assessments broadened its reach, helping screened hybrid calculations become an established part of materials research.\cite{ref044,ref045,ref034}

The impact extends beyond predicting host band gaps. HSE and its tuned variants have been used to investigate defect behavior, predict redox energetics, and identify candidate solid-state quantum emitters.\cite{ref050,ref058,ref051} Their adoption is also visible in the infrastructure of materials discovery. The Materials Project's MOF Explorer provides HSE06 band gaps for a subset of the QMOF database and identifies them as likely to be more accurate than its PBE results.\cite{mpMOFHSE} Automated HSE06 band-structure workflows are also available in the Materials Project's atomate2.\cite{mpAtomate2HSE} These uses connect the original effort to make exact exchange affordable with the practical task of selecting and understanding materials.

HSE now also serves as a reference against which new semilocal approximations are judged, while dielectric-dependent and tuned hybrids extend the treatment of exchange and screening beyond its fixed profile.\cite{ref072,ref081,ref088} These advances draw on several conceptual traditions. Together with HSE's continuing use, they show why the practical value of controlling where exact exchange acts extends beyond the particular compromises embodied in HSE03 and HSE06.

About a quarter century after its conception, HSE's significance lies in this continuity between a physical idea, a working method, and the research it enabled. It helped turn quantum-mechanical equations into practical predictions for real materials. Its enduring contribution lies both in the method itself and in the wider range of scientific questions that researchers can address with it.

\section*{Acknowledgments}

I thank Jochen Heyd, Matthias Ernzerhof, and Andreas Savin for the work and the conversations that led to HSE; Matthew C. Strain, R. Eric Stratmann, John C. Burant, John M. Millam, and Michael J. Frisch for the algorithmic foundations; and Konstantin Kudin and Artur Izmaylov for the periodic Gaussian-orbital implementation. I thank Oleg Vydrov and Artur Izmaylov for finding the screening-parameter inconsistency; Aliaksandr Krukau and Juan Peralta for the subsequent assessments; Ver\'{o}nica Barone, Jamal Uddin, and Hiromi Nakai for the early materials applications; and Ionut Prodan, Enrique Batista, Xiao-Dong Wen, Richard L. Martin, and their experimental collaborators at LANL for the actinide studies. I also thank Joachim Paier and Georg Kresse for the VASP implementation and its validation; John Perdew for two decades of collaboration; and Thomas Henderson, Benjamin Janesko, Giovanni Scalmani, Melissa Lucero, and Alejandro Garza for the later functional development, assessment, and benchmarking discussed here. G.E.S. is a Welch Foundation Chair (C-0036).

\section*{Author Declarations}

\textbf{Conflict of Interest}: The author has been a scientific collaborator of Gaussian Inc. since 1993.

\textbf{Prior publication of historical material}: Portions of the historical narrative in the first part were previously recounted in the author's scientific autobiography, Ref.~\onlinecite{ref010}, which is cited at the relevant passages. 

\textbf{Use of AI tools}: ChatGPT-6.0 Astra from OpenAI and Claude Opus 5 from Anthropic were used under the author’s direction for literature searches, synthesis of published results, drafting, and critical assessment of this Perspective. These tools helped organize the historical literature and cross-check comparisons across published benchmarks, including assembling the Tables and generating Fig. 1 from published functional definitions. No electronic-structure calculations were performed by the AI agents. The author takes full responsibility for the content, including verification of scientific and historical assertions against the primary literature.

\section*{References}
\input{bib.tex}

\end{document}

%% file: bib.tex
% References with article titles and source metadata.
% LaTeX thebibliography, included with \input{bib}; no BibTeX run required.
% Ordered by first citation in main.tex; retain each citation key when editing.